\documentclass[sigplan,screen, 10pt]{acmart}

\acmYear{2023}\copyrightyear{2023}
\setcopyright{rightsretained}
\acmConference[SOSP '23]{ACM SIGOPS 29th Symposium on Operating Systems Principles}{October 23--26, 2023}{Koblenz, Germany}
\acmBooktitle{ACM SIGOPS 29th Symposium on Operating Systems Principles (SOSP '23), October 23--26, 2023, Koblenz, Germany}
\acmPrice{15.00}
\acmDOI{10.1145/3600006.3613166}
\acmISBN{979-8-4007-0229-7/23/10}

\usepackage{tikz}
\usepackage{amsmath}
\usepackage{graphicx}

\usepackage{xspace}
\usepackage{paralist}
\usepackage{soul}

\usepackage{algpseudocode}
\usepackage[ruled,linesnumbered,vlined]{algorithm2e}
\usepackage{multirow}
\usepackage{makecell}
\usepackage{hyperref}

\hypersetup{
    colorlinks=true,
    linkcolor=blue,
    filecolor=magenta,      
    urlcolor=blue,
}

\newcommand{\OurSys}{\textsc{SPFresh}\xspace}
\newcommand{\uproto}{\textsc{LIRE}\xspace}

\newcommand{\shuotao}[1]{\textcolor{red}{(#1 --shuotao)}\xspace}
\newcommand{\cheng}[1]{\textcolor{red}{(#1 --CL)}\xspace}

\newcommand{\yuming}[1]{\textcolor{blue}{(#1 --yuming)}\xspace}
\newcommand{\fan}[1]{\textcolor{red}{(#1 --Fan)}\xspace}

\newcommand{\updater}{\emph{In-place Updater}\xspace}
\newcommand{\rebuilder}{\emph{Local Rebuilder}\xspace}
\newcommand{\blockctrl}{\emph{Block Controller}\xspace}

\newcommand{\mysubsection}[1]{\vspace{2pt}\noindent\textit{\textbf{#1}:}\xspace}

\newcommand{\revisionstart}{\color{black}}
\newcommand{\revisionend}{\color{black}}

\newcommand{\revisionsecondstart}{\color{black}}
\newcommand{\revisionsecondend}{\color{black}}

\begin{document}

\title{\OurSys: Incremental In-Place Update for Billion-Scale Vector Search }

\author{\textbf{Yuming~Xu\textsuperscript{1, 2}
\enskip
Hengyu~Liang\textsuperscript{1, 2}
\enskip
Jin~Li\textsuperscript{3, 1, 2}\footnotemark[1]
\enskip
Shuotao~Xu\textsuperscript{2}
\enskip
Qi~Chen\textsuperscript{2}
\enskip
Qianxi~Zhang\textsuperscript{2}
\enskip
Cheng~Li\textsuperscript{1}
\enskip
Ziyue~Yang\textsuperscript{2}
\enskip
Fan~Yang\textsuperscript{2}
\enskip
Yuqing~Yang\textsuperscript{2}
\enskip
Peng~Cheng\textsuperscript{2}
\enskip
Mao~Yang\textsuperscript{2}
}
}
\affiliation{
\institution{
\textsuperscript{1} University of Science and Technology of China \quad
\textsuperscript{2} Microsoft Research Asia \quad
\textsuperscript{3} Harvard University
}
\country{}
}




\renewcommand{\shortauthors}{Xu et al.}

\begin{abstract}

Approximate Nearest Neighbor Search (ANNS) on high dimensional vector data is now widely used in various applications, including information retrieval, question answering, and recommendation. 
As the amount of vector data grows continuously, it becomes important to support updates to vector index, the enabling technique that allows for efficient and accurate ANNS on vectors.

Because of the curse of high dimensionality, it is often costly to identify the right neighbors of a new vector, a necessary process for index update. 
To amortize update costs, existing systems maintain a secondary index to accumulate updates, which are merged with the main index by globally rebuilding the entire index periodically.
However, this approach has high fluctuations of search latency and accuracy, not to mention that it requires substantial resources and is extremely time-consuming to rebuild.

We introduce \OurSys{}, a system that supports in-place vector updates. At the heart of \OurSys{} is \uproto{}, a lightweight incremental rebalancing protocol to split vector partitions and reassign vectors in the nearby partitions to adapt to data distribution shifts. \uproto{} achieves low-overhead vector updates by only reassigning vectors at the boundary between partitions, where in a high-quality vector index the amount of such vectors is deemed small.
With \uproto{}, \OurSys{} provides superior query latency and accuracy to solutions based on global rebuild, with only 1\% of DRAM and less than 10\% cores needed at the peak compared to the state-of-the-art, in a billion scale disk-based vector index with a 1\% of daily vector update rate.

\end{abstract}

\ccsdesc[500]{Information systems~Information storage systems}
\ccsdesc[500]{Information systems~Information retrieval}

\keywords{Vector Search, Incremental Update, Billion-scale}




\renewcommand{\authors}{Yuming~Xu, Hengyu~Liang, Jin~Li, Shuotao~Xu, Qi~Chen, Qianxi~Zhang, Cheng~Li, Ziyue~Yang, Fan~Yang, Yuqing~Yang, Peng~Cheng, Mao~Yang}
\settopmatter{printacmref=true}
\maketitle

\footnotetext[1]{Work done during his final-year study at USTC and his internship at MSRA.}

\section{Introduction}

Today deep learning models can embed almost all types of data, including speech, vision, and text information, into \emph{multi-dimensional vectors} with tens or even hundreds of dimensions. Such vectors are critical for complex semantic understanding tasks~\cite{mikolov2013efficient,pennington2014glove}.
To enable effective vector analysis, vector nearest neighbor search (NNS) systems have become critical system components for an increasing number of online services like search~\cite{Taobao} and recommendation~\cite{Suchal2010Full}.

To satisfy the strict query latency requirement for these online services, vector search systems often resort to \emph{approximate nearest neighbor search} (ANNS)~\cite{nara2014Streaming, ADBV2020VLDB,DISKANN2019NEURIPS,MILVUS2021SIGMOD,HMANN2020NEURIPS,  VEARCH2018MIDDLEWARE, FAISS2020, manu}, to locate as many correct results as possible (i.e., query accuracy). 
At the heart of a large-scale NNS system is a \emph{vector index}, a key data structure that organizes high-dimensional vectors efficiently for high-accuracy low-latency vector searches~\cite{HDINDEX2018VLDB, SRS2014VLDB, Datar:2004:LHS, HNSW2018IEEE, KDTREE1975ACM, ChenW21}.

Like traditional indices, a high-quality vector index organizes a quick ``navigation map'' of vectors based on the vector proximity in a high dimensional space.
The proximity measurements are often implemented with ``shortcuts'', which only exist between a pair of vectors with a short distance.
A search query traverses the datasets based on ``shortcuts'' 
the result set. 
The quality of the index for efficient traversal is highly dependent on the quality of shortcuts, where insufficient shortcuts miss relevant vectors, and extraneous shortcuts incur excessive traversal and storage costs.
For high-dimensional data, vector indices require careful construction to produce a sufficient amount of high-quality ``shortcuts'', often as vector partitions~\cite{ChenW21, Datar:2004:LHS} or graphs of vector data~\cite{DISKANN2019NEURIPS, MILVUS2021SIGMOD}. 

To add fuel to fire, there is a strong desire to support \emph{fresh update} of vector indices because current systems generate a vast amount of vector data continuously in various settings.
For example, 500+ hours of content are uploaded to YouTube\cite{Youtube} every minute, one billion new images are updated in JD.com every day~\cite{VEARCH2018MIDDLEWARE}, and 500PB fresh unstructured data are ingested to Alibaba during a shopping festival~\cite{ADBV2020VLDB}.
Fresh updates require vector indices to incorporate new vectors at unprecedented scale and speed while maintaining their high-quality to produce low query latency and high query accuracy of approximate vector searches.

However, it is non-trivial for vector indices to maintain high-quality ``shortcuts'' when updating vectors with hundreds of dimensions. 
Graph-based indices have inherent high cost to update vectors in place, because each insertion or removal of vector datum often requires examining the entire graph to update the edges in a high-dimensional space.  
One silver lining to fast vector index update is that \revisionstart cluster\revisionend-based index, which is less costly to update than the graph-based index.
Vector insertion and removal only require constant local modification to vector partitions(s). 
Nevertheless, as updates accumulate per vector partition, the index quality deteriorates because the data distribution skews over time, which makes partition sizes uneven and hence hurts both query latency and accuracy~\cite{VEARCH2018MIDDLEWARE}.

Because of the difficulty of running vector index updates in place, existing ANNS system support vector updates~\cite{VEARCH2018MIDDLEWARE,nara2014Streaming,FreshDiskANN2021,MILVUS2021SIGMOD} out-of-place, by periodic \emph{rebuilding of global index}. A batch of vector updates are accumulated and indexed separately, i.e., out-of-place, and are periodically merged to the base index by rebuilding the entire index. 
Such practices introduce significant resource overheads to ANNS systems.
For example, to build a global DiskANN index for a 128G SIFT dataset with a scale of 1 billion, it would require a peak memory usage of 1100GB for 2 days, or 5 days under a memory usage of 64GB with 32 vCPUs~\cite{DISKANN2019NEURIPS}. Such rebuilding can even consume more resources than the index serving costs (\S\ref{sect:bg:challenges}).
In addition, such an out-of-place update method hurts the search performance of online services because it follows a Log-Structured-Merge (LSM) style for updates~\cite{FreshDiskANN2021}, which trades read performance for write optimization.


To scale to large vector datasets with lower costs, this paper presents \OurSys{}, a disk-based vector index that supports lightweight incremental in-place local updates without the need for global rebuild. 
\OurSys{} is based on the state-of-the-art \revisionstart cluster\revisionend-based vector index design, capable of incorporating vector index updates online with low overheads while maintaining good index quality for high search performance and accuracy for billion-scale vector datasets.
The core of \OurSys{} is \uproto{}, a \underline{L}ightweight \underline{I}ncremental \underline{RE}-balancing protocol that accumulates small vector updates of local \revisionstart vector \revisionend partitions, and re-balances local data distribution at a low cost.
Unlike expensive global rebuilds, LIRE is capable of maintaining index quality by fixing data distribution abnormalities locally on-the-fly.

The key design rationale behind \uproto{} is to leverage a vector index that is \emph{already} in a well-partitioned state. Small vector updates to a high-quality vector partition may only incur changes in itself and its neighboring partitions. Because the updates are small, the corresponding changes are most likely to be limited in a local region. This makes the entire rebalancing process lightweight and affordable.

Despite this opportunity, rebalancing is still non-trivial. 
In particular, LIRE needs to address the following challenges.
1) In order to keep search latency short, LIRE needs to maintain an even distribution of partition sizes via timely split and merge.
2) In order to keep search accuracy high, LIRE needs to identify the smallest set of vectors that cause data imbalance in the index. These vectors should be reassigned to maintain high index quality.
3) An implementation of LIRE should be lightweight with negligible performance impacts on the foreground search.

LIRE tackles these challenges by making the following four contributions: 
\begin{compactitem}[\noindent$\bullet$]
\item LIRE keeps partition size distribution uniform by splitting and merging partitions proactively and incrementally. 
\item LIRE formally identifies two necessary conditions for vector reassignment based on the rule of nearest neighbor posting assignment (NPA). With the necessary conditions, LIRE opportunistically identifies a minimal set of neighborhood vectors to adapt to data distribution shifts. 
\item An implementation of LIRE is decoupled as a two-stage feed-forward pipeline, which moves the background split-reassign off from the critical path of foreground update. Each pipeline stage is multi-threaded to saturate the high IOPS of a high-performance NVMe device.
\item An SSD-backed user-space storage engine dedicated to LIRE, which bypasses legacy storage stack, prioritizes partition reads, and optimizes for partition appends.
\end{compactitem}

Experiments show \OurSys{} outperforms state-of-the-art ANNS systems that support fresh updates on all fronts, with low and constant search/insert latency, high query accuracy, as well as efficient resource usages for billion-scale vector datasets. Instead of an additional 1000GB memory and 32 cores needed by DiskANN global rebuild, \OurSys{} outperforms DiskANN by 2.41$\times$ lower tail latency on average with only 
10GB memory and 2 cores. Moreover, \OurSys{} reaches the IOPS limitation with stable performance and resource utilization. It simultaneously reaches peak 4K QPS search throughput and 2K QPS update throughput on a single NVMe SSD disk with 15 cores.

\section{Background and Related Work}
\label{sect:bg}

\revisionstart
In this section, we present the basic operations in ANNS-based vector search and introduce two mainstream on-disk vector indices and their respective index update challenges.
\revisionend

\subsection{Vector Search and ANNS}
\begin{figure}[!t]
\includegraphics[width=\columnwidth]{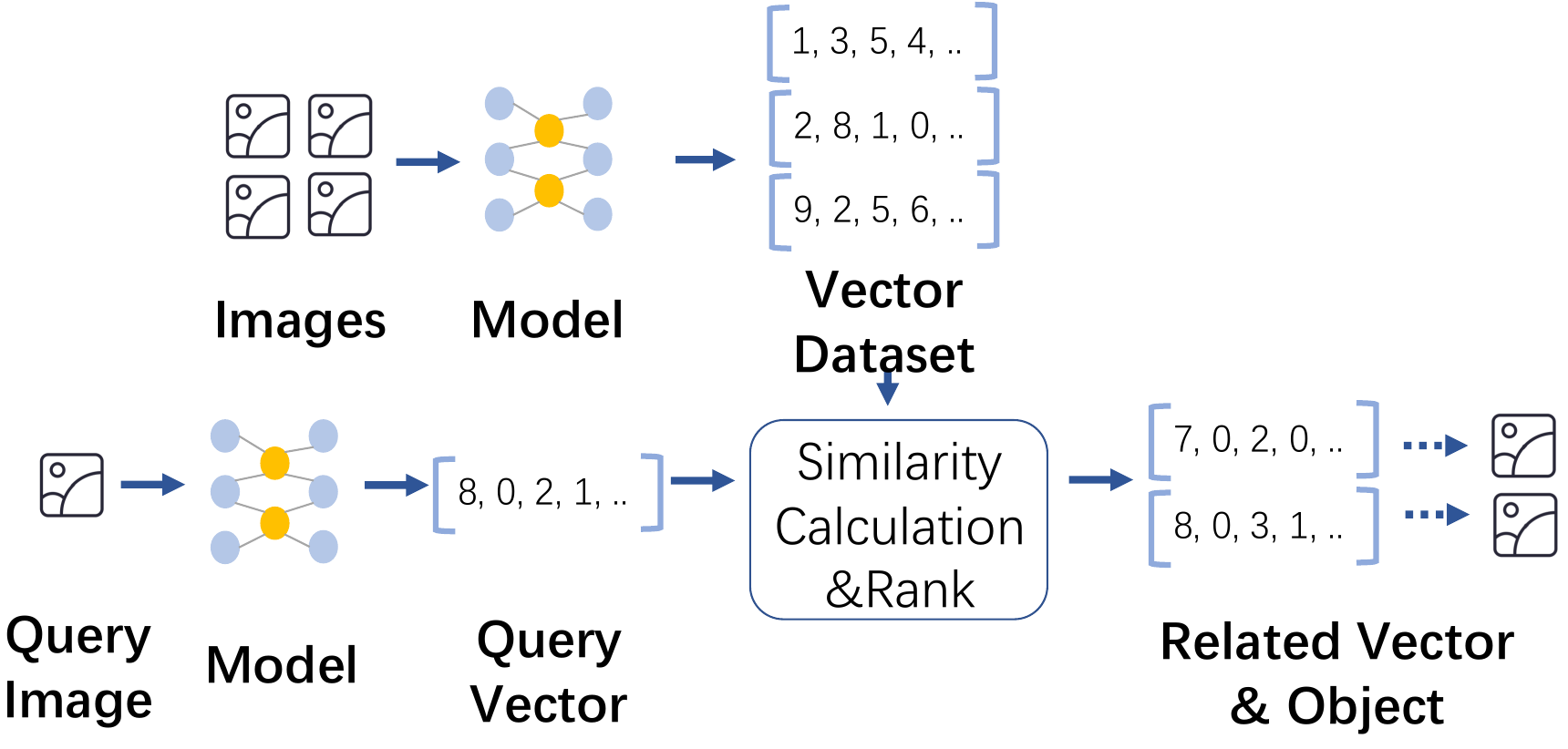}
\caption{\revisionsecondstart{An example of vector search: to find the most similar images in an image dataset. Query images and data images are all represented as vectors.}\revisionsecondend}
\label{fig:vectorsearchexample}
\end{figure}

\revisionsecondstart
A common use-case of vector search involves finding the most similar items in a large dataset based on a given query. This process is often used in recommendation systems, search engines, and natural language processing tasks.
As shown in Figure \ref{fig:vectorsearchexample}, to find similar images from dataset given a query image, the system first represents each image in the dataset as a high-dimensional vector through a deep learning model. The query image is also encoded into a vector in the same high-dimensional space. Then, the system calculates the similarity between the query vector and each vector in the dataset using a similarity metric, such as cosine similarity or Euclidean distance. The system ranks the images based on their similarity scores and returns the top results to the user. Essentially, the search is to find the query vector's nearest neighbors in a high-dimensional space.
\revisionsecondend
\label{sect:bg:ANNS}

Formally, 
given a vector set $X \in \mathbf{R}^{n \times m}$ containing $n$ $m$-dimensional vectors and a query vector $\mathbf{q}$, vector nearest neighbor search aims to find a vector $\mathbf{x}^*$ from $X$ such that $\mathbf{x}^*= \arg\min_{\mathbf{x} \in \mathbf{X}}\operatorname{Dist}(\mathbf{x}, \mathbf{q})$, where $\operatorname{Dist}$ is the similarity metric discussed above.
This definition can be extended to $K$-nearest neighbor (KNN) search~\cite{ChenW21}. 
Modern machine learning models typically generate vectors with dimensions ranging from 100 to 10,000, or even more.
For example, GPT-3 generates four sizes of embedding vectors with dimensions ranging from 1024 to 12288~\cite{GPT3embedding}. The high dimensionality makes it challenging to find the exact $K$ nearest neighbors efficiently~\cite{clarkson1994algorithm}. To address this issue, recent systems commonly rely on \emph{approximate} nearest neighbor search (ANNS)~\cite{DISKANN2019NEURIPS,HMANN2020NEURIPS,ChenW21} to make the effective trade-off across resource cost, result quality, and search latency, thus scaling to large vector datasets. 

\revisionstart
Due to its approximate nature, search result accuracy
becomes an important metric to gauge the quality of a vector index. 
In ANNS, \verb|RecallK@K| is commonly used to measure result quality. 
For an approximate KNN query, \verb|RecallK@K| is defined as $\frac{|Y\cap G|}{|G|}$, where $Y$ is the query's result set, and $G$ is the query's ground truth result set, $|Y|=|G|=K$.

\revisionend

\revisionstart
\subsection{Vector Index Organization}

A vector index can be abstracted as a logical graph, where a vertex represents a vector, and an edge denotes the close proximity of two vectors in terms of distance.
And vector indices for ANNS can be categorized into \emph{fine-grained graph-based vector indices} and \emph{coarse-grained cluster-based vector indices}. These two methods can be applied to both in-memory or on-disk scenarios.


In this paper, we only focus on on-disk vector indices since they are more cost-effective for large-scale vector-sets. 
Meanwhile, they pose a unique challenge for vector updates since disk writes are much more 
costlier than DRAM writes.

\noindent\textbf{\textit{Fine-grained graph-based vector indices}} represent each vector as a vertex,  
and an edge exists between two vertices if they are close in distance. 
Locating $K$ nearest vectors often involves best-first graph traversals, where neighboring vertices are explored in ascending distance order.

Example vector index solutions based on fine-grained graphs include \emph{neighborhood-graph based methods}~\cite{HajebiASZ11,DongCL11,WangWZTGL12,zbMATH03016052,toussaint1980relative,gabriel1969new,FuNSG17,HNSW2018IEEE} which organize all the vectors into a neighborhood graph with each vector connected to its nearest vectors, and \emph{hybrid methods}~\cite{wang2012query, SPTAG2018, NGTONNG2018, NGTPANNG2016} which consist of space-partition trees and a neighborhood graph to take advantage of both tree and graph data structures. \emph{Space-partition-tree based methods}~\cite{BentleyCACM1975,FreidmanTOMS1977,BeisCVPR1997,SproullALGO1991,YianilosSODA1993,AryaJACM1998,MooreUAI2000,LiuNIPS2004,NisterCVPR2006,DasguptaSTOC2008,McCartin-LimICML12,WangPAMI2014,MujaPAMI2014,BabenkoCVPR17} can be treated as a special kind of fine-grained graphs. They use a tree to represent the space division and the vector to subspaces mapping.
Most of these solutions are based on in-memory implementations for performance and are expensive to scale to billion-scale data-sets.

Only a few fine-grained graph-based vector indices are optimized for secondary storage (e.g., DiskANN~\cite{DISKANN2019NEURIPS} and HM-ANN~\cite{HMANN2020NEURIPS}).
Similar to external graph systems~\cite{fast15-flashgraph,osdi12-graphichi,sosp13-xtream, isca18-grafboost}, these fine-grained graph-based vector indices are stored in two parts: vertex data and edge data as vertex adjacency lists.
Edge data are stored in secondary storage, and vertex data are either on disk~\cite{HMANN2020NEURIPS} or in memory~\cite{DISKANN2019NEURIPS}, where in-memory vertex data speed up computation, i.e. distance calculation in the case of ANNS~\cite{DISKANN2019NEURIPS}.

To reduce search costs, DiskANN~\cite{DISKANN2019NEURIPS} employs a fixed graph traversal strategy, where it caches the neighborhood of the fixed starting point \emph{in memory} to speed up graph traversal in the initial stage.
DiskANN further maintains an in-memory copy of compressed vertex data (using product quantization) to speed up distance calculation during graph traversals.
In contrast, HM-ANN~\cite{HMANN2020NEURIPS} constructs a hierarchical in-memory graph where it can 
navigate to the nearest entry point to the main graph on secondary storage, and thus efficiently identify the target region for nearest vectors.

Although effective for vector search, graph-based vertex indices are unfriendly to updates (details in \S\ref{sect:bg:challenges}). 



\noindent\textbf{\textit{Coarse-grained cluster-based vector indices}}
organize vector indices via clustering, where vectors in close proximity are kept in the same partition. 
Logically, vectors in each partition represent a fully-connected graph, while vectors across different partitions have no edge.
Since no explicit edge data are required, coarse-grained cluster-based vector indices require much smaller storage.
Vector search on cluster-based vector indices first identifies candidate partitions by measuring the distance to the partitions' centroids and then calculates the $K$ nearest vectors from the candidate partitions via a full scan.

Coarse-grained cluster-based vector indices include \emph{hash-based methods}~\cite{Datar:2004:LHS,He:2010:SSS,Jain:2008:Learned,kulis2009kernelized,raginsky2009locality,wang2010semi,mu2010non,song2011multiple,weiss2009spectral,xu2011complementary} which use multiple locality-preserved hash functions to do the vector-to-partition mapping, and \emph{quantization-based methods}~\cite{jegou2011searching,baranchuk2018revisiting,zhang2019grip,babenko2014inverted,FAISS2020} which use Product Quantization(PQ)~\cite{jegou2010product} to compress the vectors and KMeans to generate the vector-to-partition mapping codebooks.


A cluster-based vector index should preserve the balance across partitions to achieve low tail search latency. 
However, ANNS indices leveraging locality-sensitive hashing~\cite{Datar:2004:LHS, Jain:2008:Learned, weiss2009spectral, xu2011complementary, WangZSSS18} and k-means~\cite{kmeans} for clustering pay less attention to partition balance. 
Such ANNS indices often produce uneven partitions and thus are only adopted by in-memory systems where the absolute tail latency is much less pronounced than that of an on-disk solution. 

SPANN~\cite{ChenW21} is the first on-disk vector index that achieves low tail search latency through balanced clustering.
SPANN divides a vector-set into a large number of balanced partitions stored on disk and keeps the centroids of the partitions in the memory for quick identification of candidate partitions during search. 
It employs several techniques to ensure a well-balanced partition state (details in \S\ref{sect:design:SPANN}).
SPANN achieves state-of-the-art performance on memory cost, result quality, and search latency across multiple billion-scale datasets.

Cluster-based vector indices are \emph{friendly to updates} because each vector insertion or deletion only involves local modifications of vector data in the corresponding partition. 
However, a naive update on local partitions may eventually lead to imbalanced clusters and consequently deteriorate search tail latency and accuracy (more in \S\ref{sect:bg:challenges}). 

\revisionend

\subsection{Freshness Demands and Challenges}
\label{sect:bg:challenges}

Modern ANNS systems are required to accommodate billions of vector updates every day while still preserving low query latency and high query accuracy. With the new popular OpenAI ChatGPT retrieval plugin~\cite{chatgptplugin}, some AI applications built atop even require real-time updates to keep up with the updates on their personal documents or contexts, such as files, notes, emails, and chat histories, all in the form of vector, in order to retrieve most relevant snippets as new prompts. 
However, it is non-trivial for vector indices to maintain index quality when updating vectors. 

\noindent\textbf{Out-of-place update.}
For vector inserts, fine-grained graph-based indices have to connect a new vector to hundreds of neighboring vectors in order to maintain sufficient shortcuts in the high-dimensional space. Deletions of vectors are even more expensive as they often involve the total scan of a unidirectional graph. 


To overcome the difficulty of in-place updates, 
existing systems resort to 
out-of-place updates with \emph{periodical global updates}. 
These systems accumulate and index delta vector updates in a separate, secondary in-memory index, which is periodically merged to the base index by a global index rebuilding process to maintain good index quality. 
Many popular ANNS systems, such as ADBV~\cite{ADBV2020VLDB} and Milvus~\cite{MILVUS2021SIGMOD}, use this method. To defer expensive global updates, Milvus even introduces multiple delta indices in memory.
However, this approach requires vector search to examine both main and secondary indices, which increases resource demands and hurts search performance.
Table~\ref{tbl:rebuild_cost} shows global rebuilds are both resource-hungry and time-consuming. For example, rebuilding a 1-billion vector index~\cite{DISKANN2019NEURIPS} for DiskANN, a recent disk-based system, needs 1100GB DRAM, 32 vCPUs, for 2 days. When limiting resources to 64GB memory and 16 vCPUs, the rebuilding time becomes significantly longer, e.g., 5 days for DiskANN. This stressful setting could also lead to a catastrophic drop in query performance because of severe computational resource starvation.

\begin{table}[t!]
    \centering
    \small
    \begin{tabular}{|c|c|c|c|c|}
    \hline
     & Memory & CPU & Time \\
     \hline
      DiskANN & \makecell{1100 GB \\ 64GB} & \makecell{32 cores \\ 16 cores} & \makecell {2 days \\ 5 days} \\
     \hline
     SPANN & 260 GB & 45 cores & 4 days \\
     \hline
    \end{tabular}
    \caption{Global rebuild costs of disk-based ANNS indices for billion-scale datasets.}
    \label{tbl:rebuild_cost}
\end{table}

\begin{figure}[t!]
\includegraphics[scale=1]{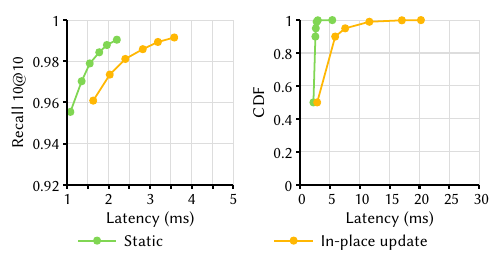}
\caption{Recall and tail latency in two system settings, namely, \textit{static} and \textit{in-place update}. The static setting refers to an index of 2 million vectors, while the in-place update setting refers to an index built by \revisionstart applying 0.5 million vector updates to a base index of 1.5 million vectors. \revisionend 
The index system is SPANN, and the dataset is sift~\cite{Sift}. 
}
\label{fig:Motivation}
\end{figure}

\noindent\textbf{Early attempts to in-place update.}
\revisionstart Compared to out-of-place vector updates, few systems support in-place updates. Vearch~\cite{VEARCH2018MIDDLEWARE} is one of such systems based on cluster-based in-memory vector indices, where it inserts a new vector to its nearest partition (a.k.a. posting, the partition is implemented as a posting list) \revisionend and supports deletions by maintaining a tombstone bitmap for result filtering. 

To understand the impact of Vearch's design to on-disk index, we apply Vearch's design to SPANN, the \emph{only} partition-based on-disk vector index system. 
Figure \ref{fig:Motivation} shows that 
\revisionstart updating one-third of the vectors degrades the query recall by more than one point and increases tail latency by 4X, \revisionend compared to static index building. The reasons are two-fold:
1) With the growth of the data size, query latency will increase due to the expansion of the posting length.
2) Since the centroids for each partition are fixed, the recall will decline as static centroids cannot capture the gradual distribution shift in the partition.
To conclude, to maintain high index quality and stabilize search latency, although Vearch and the modified SPANN do not require \emph{out-of-place} data structure, they still require periodical global rebuilds. For instance, Vearch performs weekly rebuilds.
The rebuild overheads might be acceptable for \emph{in-memory} vector indices.
However, for disk-based indices like SPANN, global rebuilds are expensive, as shown in Table~\ref{tbl:rebuild_cost}. 

\revisionstart 
In summary, existing graph-based and cluster-based solutions, regardless of in-place or out-of-place, all rely on periodic global index rebuilding to preserve index quality and stabilize search performance. However, this process entails considerable resource consumption. \revisionend

\subsection{Our Goals}
To this date, efficient fresh update for disk-based vector index is still an open challenge. 
In this paper, we aim to propose a new disk-based ANNS system to fulfill the following goals: 1) low resource cost to maintain the index for large-scale vector datasets; 2) support high throughput and low latency vector queries for both search and update; and 3) new vectors can be recalled in high probability. 

To achieve this, motivated by the above understandings, we choose to follow the \revisionstart coarse-grained cluster-based \revisionend approach to build our on-disk index, but differ from existing solutions significantly by avoiding global rebuilds completely. 
\revisionstart The proposed solution, \OurSys, performs in-place, incremental updates in the index data structure to adapt to the data distribution shift. To this end, \revisionend \OurSys{} incorporates a \underline{L}ightweight \underline{I}ncremental \underline{RE}-balancing (LIRE) protocol, which efficiently identifies a minimal amount of partition updates introduced by new vectors for maintaining index property and thus eliminating visible accuracy loss. Equally importantly, we also address a few system challenges to make LIRE re-balance sufficiently fast and cheap to alleviate negative impacts on search latency, in particular, \emph{tail latency}.  \revisionstart Essentially, LIRE can be considered as an efficient compaction technique in the high-dimensional space. \revisionend

\section{LIRE Protocol Design}
\label{sect:design:LIRE}

\uproto{} is built on SPANN~\cite{ChenW21}, the state-of-the-art disk-based vector index system. In this section, we first introduce SPANN briefly and then elaborate \uproto{} in detail.

\revisionstart 
\subsection{SPANN: A Balanced Cluster-based Vector Index}
\revisionend
\label{sect:design:SPANN}

\begin{figure}[t!]
\includegraphics[width=\columnwidth]{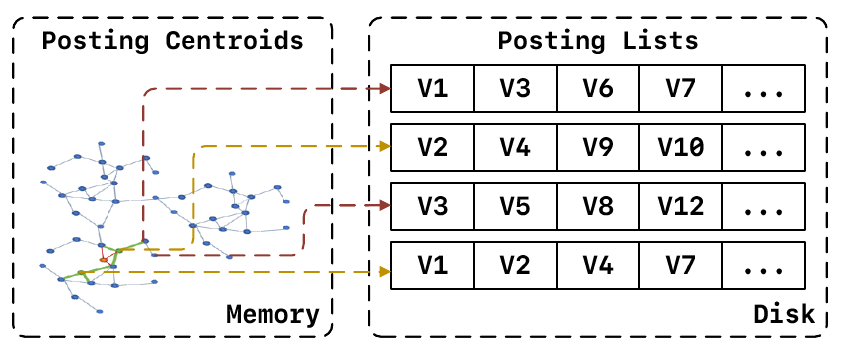}
\caption{SPANN index data architecture.}
\label{fig:spannindex}
\end{figure}

\revisionstart
SPANN~\cite{ChenW21} is a billion-scale cluster-based vector index optimized for secondary storage. 
\autoref{fig:spannindex} shows the overall SPANN index structure.
SPANN stores the vectors as a large number of postings\footnote{We use ``posting'' and ``partition'' interchangeably in this paper.} on disk, each represents a cluster of close-by vectors. 
Moreover, SPANN organizes a graph-based in-memory index, SPTAG~\cite{SPTAG2018}, for the centroids of all postings, to quickly identify relevant postings for a query. 
\revisionend

For a query, it first identifies the closest posting centroids through the in-memory index, and then loads the corresponding postings from disk to memory for further search. 
\revisionstart
To control the tail latency and maintain high search recall, SPANN makes postings \emph{well balanced} by maintaining two key properties.
1) SPANN divides the vectors \emph{evenly} into a large number of small-sized postings by a fast hierarchical balanced clustering algorithm, so that each query visits a similar amount of vectors for bounded search tail latency. 
2) SPANN replicates a few vectors in boundaries across postings, which sufficiently maintains high search recalls.
\revisionend

The \emph{balanced} SPANN index inspires us to propose a new lightweight incremental re-balancing (LIRE) protocol. \textit{The intuition here is a single vector update to a well-balanced index may only incur changes in a local region.}  This makes the entire rebalancing process lightweight and affordable.

\subsection{LIRE: \underline{L}ightweight \underline{I}ncremental \underline{RE}-balancing}
\label{ssec:lire}


A key property of a well-partitioned vector index is the \emph{nearest partition assignment} (NPA): each vector should be put into the nearest \revisionstart posting \revisionend so that it can be well represented by the \revisionstart posting \revisionend centroid. 
As continuous vector updates to a \revisionstart posting \revisionend may degrade query recalls and latency, \OurSys{} will split a \revisionstart posting \revisionend after it grows to the preset maximum length. However, a naive splitting can violate the NPA property of the index. 

\begin{figure}[!t]
\centering
\includegraphics[scale=0.5]{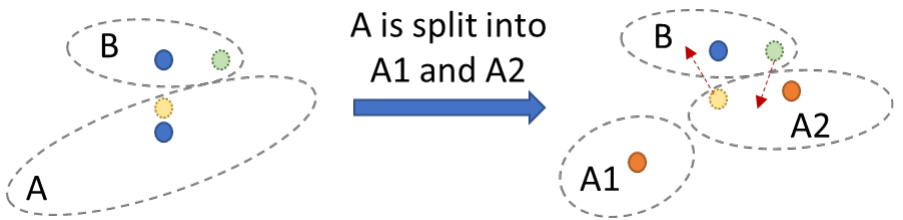}
\caption{\revisionstart Posting \revisionend split violates the 
NPA property.}
\label{fig:challenge_example1}
\end{figure}

Figure~\ref{fig:challenge_example1} illustrates a case of NPA violation. 
Originally, there were two \revisionstart postings, \revisionend A and B, near each other, where the blue dots represent their centroids. 
At a certain point, \revisionstart posting A exceeds the length limit \revisionend upon vector insertions and is split into two new \revisionstart postings, \revisionend A1 and A2. 
The orange dots represent the new centroids of A1 and A2 elected after the split. 
With a naive split, the vectors in \revisionstart posting \revisionend A will \revisionstart \emph{only} \revisionend go to \revisionstart Posting \revisionend A1 and A2 respectively, \revisionstart based to their distance to new centroids. \revisionend
For illustrative purposes, we assume the yellow dot (a vector) goes to A2. 

\revisionstart However, the creation of new centroids via a spit makes previous NPA-compliance obsolete for vectors in the nearby postings, A1, A2 and B.
\revisionend
First, the nearest \revisionstart posting \revisionend of the yellow dot \revisionstart changes to \revisionend B, since B's centroid is closer to the yellow dot than A2's centroid. In this case, using the centroid of A2 to represent the yellow dot violates NPA. 
Second, the nearest \revisionstart posting \revisionend of the green dot, which was B \revisionstart before the split, changes to A2. \revisionend
These two violations degrade the index quality and result in low recalls. 

\revisionstart To fix the NPA violations after splits and maintain the high index quality, we design \uproto{} protocol, which reassigns vectors in nearby postings of a split.
At its core, \uproto{} protocol consists of five basic operations: Insert, Delete, Merge, Split, and Reassign.
\revisionend

\mysubsection{Insert \& Delete} LIRE directly inserts a new vector to the nearest partition following the original SPANN index design. 
\uproto{} also ensures the deleted vectors will not appear in the search results and will eventually be removed from the corresponding postings. 

Note that Insert and Delete are \emph{external interfaces} exposed to users.
The remaining three operations are \emph{internal interfaces} and thus are oblivious to users. 
These three operations work together to keep the size of the posting small and balanced and to ensure vectors are assigned to the right posting, following the NPA property.


\mysubsection{Split} When a posting exceeds a length limit, \uproto{} evenly splits the oversized posting into two smaller ones. 
As introduced in the previous section, vectors in the neighboring postings may violate the NPA property after the split.
Thus, a reassign process \revisionstart (detailed in \S\ref{ssec:reassign}) \revisionend will be triggered for the vectors in the split postings as well as nearby postings.

\mysubsection{Merge} When a posting size is \revisionstart smaller than a lower threshold, \uproto{} identifies its nearest posting as candidates for merging. \revisionend
\revisionstart In particular, \uproto{}'s merge process deletes one posting with its centroid (e.g., the shorter posting), and appends them to the other posting directly.
\revisionstart
After that, a reassign process is required for the vectors of the deleted posting because the deletion of their old centroid might break the NPA rule after being merged with the other posting.
\revisionend
\revisionstart
Reassignments will not induce splits of the merged posting because vectors can only be reassigned out. However, a reassigned vector may trigger the split to the target posting. Despite such cascading effects, \S\ref{ssec:converge} shows that LIRE's split-reassignment process will always converge.
\revisionend



\subsection{Reassigning Vectors} 
\label{ssec:reassign}

\revisionstart
Reassigning vectors can be expensive, because they require expensive changes to the on-disk postings for each reassigned vector. 
\revisionend
Thus it is critical to identify the right set of neighborhoods (neighboring postings) to avoid unnecessary reassignment. 
For a merged \revisionstart posting \revisionend, only vectors from \revisionstart deleted posting \revisionend require reassignment, \revisionstart because the deletion of a centroid does not break NPA compliance of vectors from undeleted postings. \revisionend

\revisionstart
On the other hand, a split not only deletes a centroid but creates two new ones.
Therefore a split creates more complex scenarios of potential NPA violations.
After examining \autoref{fig:challenge_example1}, we derive two necessary conditions for reassignment after splitting, assuming the high-dimensional vector space is Euclidean. 
\revisionend

First, a vector $v$ in the old posting with centroid $A_o$ is required to consider being reassigned if:
\begin{equation}
    D(v, A_o) \leq D(v, A_i), \forall i \in {1, 2} \label{eq:ne1}
\end{equation}
where $D$ denotes the distance, $A_o$ represents the old centroid before splitting,  and $A_i$ represents any of the two new centroids. 
This reassignment condition means that if the old (deleted) centroid $A_o$ is the closest centroid to the vector $v$, compared to new centroids ($A_1$ and $A_2$), then it cannot be ruled out the possibility that $v$ is closer to a centroid of some nearby posting than new centroids (e.g., $B$ in \autoref{fig:challenge_example1}). 
Note that this is a \emph{necessary condition}. 
On the contrary, if $D(v, A_o) > D(v, A_i)$, this shows $v$ is having a better centroid than the old one. In this case, the neighboring certroid (e.g., $B$) cannot be better than the new ones, i.e., $D(v, B) > D(v, A_i)$ based on the NPA property of $D(v, B) > D(v, A_o)$. Thus there is no need to check reassignment in this case.

Second, a vector $v$ in the nearby posting with centroid $B$ needs to consider being reassigned if:
\begin{equation}
D(v, A_i) \leq D(v, A_o), \exists i \in {1, 2} \label{eq:ne2}
\end{equation}
This is a \emph{necessary condition} for a vector in posting $B$ to be reassigned to a newly split posting with centroid $A_i$. 
\autoref{eq:ne2} suggests $v$'s new neighboring centroids are getting closer (better) than the old (deleted) one. 
Therefore it is necessary to check if the new and closer centroids are in fact closer than \revisionstart $v$'s existing centroid B \revisionend (the blue dot w.r.t the green dot in \autoref{fig:challenge_example1}). 
\revisionstart
On the other hand, if the two new centroids are farther away from any vector $v$ outside the old posting, this means the two centroids are worse than the old one $A_o$, which is already farther away than $v$'s existing centroid. In this case, there is no need to check the reassignment of $v$. Hence the necessary condition.

\revisionend


According to the two necessary conditions, a \revisionstart complete \revisionend checking process \revisionstart would be \revisionend extremely expensive, \revisionstart because it requires computing and comparing $D(v, A_o)$ and $D(v, A_i), i \in {1, 2}$ for \emph{all vectors} in the dataset. \revisionend 
To minimize the cost, \uproto{} only examines nearby postings for reassignment check \revisionstart by selecting several $A_o$'s nearest postings, \revisionend
over which two condition checks were applied to generate the final reassign set. 
Experiments in Section~\ref{sect:eval} show empirically that only a small number of nearby postings for the two necessary condition checks is enough to maintain the index quality.

After obtaining vector candidates for reassignment, \uproto{} \revisionstart executes the reassignment. \revisionend
For vector candidate $v$, \uproto{} first searches $v$'s new closest \revisionstart posting \revisionend, then performs NPA check to get rid of false-positives: if \revisionstart a vector actually does not need reassignment, \revisionend the reassign operation is aborted. 
Otherwise, \uproto{} appends $v$ in the newly identified posting that is NPA-compliant and then deletes $v$ in the original posting.

\revisionstart
\subsection{Split-Reassign Convergence}
\label{ssec:converge}
In this section we prove that a split-reassign action to the vector index, despite the potential of triggering cascading split-reassign actions, will converge to a final state and terminate in finite steps. We first formally define the states of vector index and the events triggering state transitions. Then we prove that state transition will converge and terminate.

\mysubsection{Index State}
The state of a vector index for a vector data-set $V$ comprises of two parts. 
\begin{equation} C: \text{set of posting centroids.} \end{equation}
\begin{equation} M: \text{vector membership to centroid(s).} \end{equation}

\noindent
According to LIRE, given $C$, each vector in $M$ is assigned to its nearest centroid in $C$, i.e., $M$ is \emph{uniquely} determined by $C$.

\mysubsection{Index-State Transitions}
The state transition is triggered by two types of events:
\begin{equation} E_{insert}: \text{a vector $v$ is inserted into the vector index.} \end{equation}
\begin{equation} E_{delete}: \text{a vector $v$ is deleted into the vector index.} \end{equation}
A reassign of vector $v$ is considered as an $E_{delete}$ of $v$ followed by an $E_{insert}$ of $v$.
Note that an event will change the state of $M$, however it would not necessarily alter the state of $C$. 

Also note that only an event $E_{insert}$ may incur a \emph{split action} of the vector index, which alters the state of $C$ and subsequently $M$. Since $C$ uniquely determines $M$, we can only focus on the state change of $C$. 

\mysubsection{Split-Reassign Convergence Proof}
$E_{delete}$ may eventually trigger a merge during a search process (according to LIRE), and the merge obviously will terminate.

Suppose an $E_{insert}$ triggers a sequence of changes of $C$, denoted as $C_{i}$,$C_{i+1}$, ..., $C_{i+N}$. 
To prove the convergence is to show that $N$ is a finite number.

We note that $C$ has the following properties:
\begin{compactitem}[\noindent$\bullet$]
    \item $|C| \leq |V|$: The cardinality of $C$ is bounded and no greater than the cardinality of $V$, i.e., the vector dataset.
    \item $|C_{i+1}| = |C_i| + 1$: Each split action will delete an old centroid from $C_i$, and adds two new centroids to it. Therefore, the cardinality of $C$ \emph{always} increases by one per split action.
\end{compactitem}

\noindent
Based on Property 2, $|C_{i+N}| = |C_{i}| + N$. And according to Property 1, $N \leq V - |C_{i}|$ because $|C_{i+N}| \leq V$.
Since $|V|$ is finite, $N$ must also be finite.
Therefore the split action must terminate in finite steps. \qed

\revisionend

\begin{figure}[t!]
\includegraphics[width=\columnwidth]{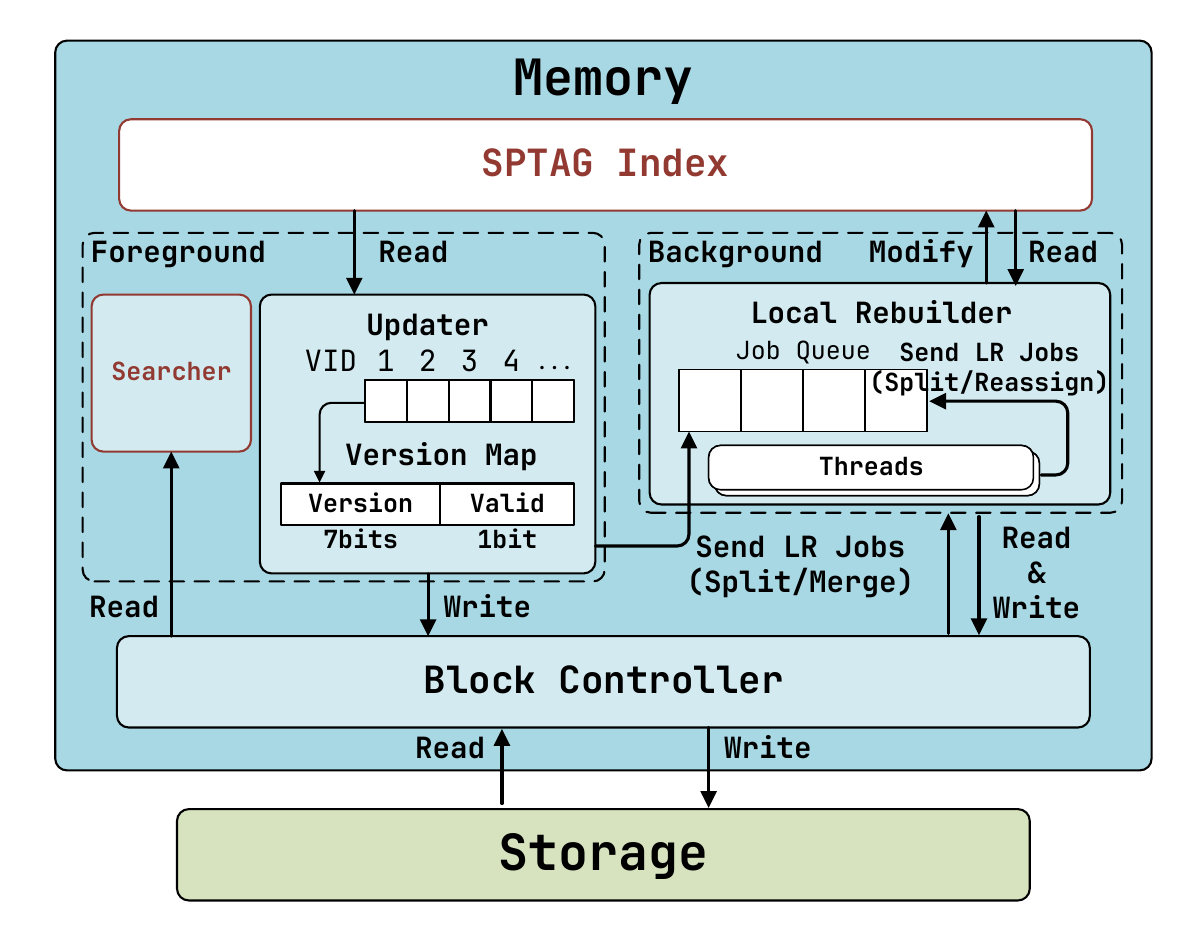}
\caption{\OurSys{} architecture (LR means Local Rebuild).}
\label{fig:architecture}
\end{figure}

\section{\OurSys\ Design and Implementation}


\subsection{Overall Architecture}
\label{sect:imple:overall}


Figure~\ref{fig:architecture} shows the system architecture of \OurSys. 
\OurSys\ reuses the SPANN SPTAG index (depicted in Figure~\ref{fig:spannindex}) for fast posting centroid navigation as well as its searcher to serve queries. It further introduces three new modules to implement \uproto{}, namely, a light-weight \updater, a low-cost \rebuilder, and a fast storage \blockctrl. 

\noindent\textit{\textbf{Updater}} appends a new vector at the tail of its nearest posting and maintains a \textit{version map} to keep track of vector deletion by setting a corresponding tombstone version to prevent deleted vectors from appearing in the search results. The map is also used to trace the replica of each vector. By increasing the version number, it marks the old replicas as deleted. \revisionstart The system keeps a global in-memory version map and stores vectors along with the version number on disk. A vector is stale if the in-memory version number is greater than that on the disk. This can be used for garbage collection caused by reassignment. The use of version can defer and batch the garbage collection so as to control the I/O overhead of vector removal. \revisionend  
After vector \revisionstart insertions \revisionend are completed in-place, the Updater checks the length of the posting and then sends a split job to \rebuilder{} if the length exceeds the split limit. 
The actual data deletions are performed asynchronously as a batch during local rebuild phase when the posting length exceeds the limit.

\noindent\textbf{\rebuilder} is the key component to implement LIRE. It maintains a job queue for split, merge, and reassign jobs and dispatches jobs to multiple background threads for concurrent execution. 
\begin{compactitem}[\noindent $\bullet$]
\item A \emph{split job} is triggered by \textit{Updater} when a posting \revisionstart exceeds the split limit. \revisionend 
It cleans deleted vectors in the \revisionstart oversized \revisionend posting and splits it into small ones if needed. 
\item A \emph{merge job} is triggered by the \textit{Searcher} \revisionstart if it finds some postings are smaller than a minimum length threshold. \revisionend
It \revisionstart merges \revisionend nearby \revisionstart undersized \revisionend postings into a single one. 
\item A \emph{reassign job} is triggered by a split or merge job, which re-balances the assignment of vectors in the nearby postings.
    
\end{compactitem}
When the background split and merge jobs are complete, \OurSys\ will update the memory SPTAG index with the new posting centroids to replace the old one.

\noindent\textbf{\blockctrl} 
serves posting read, write, and append requests, as well as posting insertion and deletion operators on disk.
It uses the raw block interface of SSD directly to avoid unnecessary read/write amplification incurred by some general storage engines, such as Log-structured-merge-tree-based KV store. 
Each posting may span multiple SSD blocks, each of which stores multiple vectors (including vector ID, version ID, and raw data). 
The \blockctrl also maintains an in-memory mapping from the posting ID to its used SSD blocks as well as the free SSD blocks pool.

Next, we will discuss the design and implementation of \rebuilder(\S\ref{sect:imple:rebuilder}) and \blockctrl(\S\ref{sect:imple:storage}) in detail. 

\subsection{Local Rebuilder Design}
\label{sect:imple:rebuilder}


In order to move split, merge, and re-assign jobs off the update critical path, \OurSys divides the update process into two parts, a foreground \textit{Updater} and a background \rebuilder. 
These two components form a feed-forward pipeline, where \textit{Updater} is the producer of requests to the \rebuilder. 
In this pipeline, the background \rebuilder is a key module that 
implements merge, split, and reassign operators of LIRE protocol efficiently to keep up with the foreground \textit{Updater}.


\subsubsection{Rebuild Operators of LIRE protocol}
\rebuilder implements LIRE with three basic operators.

\mysubsection{Merge}
To execute a merge job, the \rebuilder simply follows the merge protocol described in \S\ref{ssec:lire}.

\mysubsection{Split} 
After receiving an oversized posting split job, the \rebuilder{} first garbage collects deleted vectors in the posting and verifies whether the posting length after garbage collection still exceeds the split limit. 
If not, the \rebuilder writes the garbage-collected posting back to storage and completes the split job. 

Otherwise, a \emph{balanced clustering process} is triggered to split the oversized posting into two smaller ones.
In particular, \rebuilder leverages the multi-constraint balanced clustering algorithm in~\cite{ChenW21} to generate high-quality centroids and balanced postings.

After splitting, \rebuilder puts two new postings back to the index and deletes the original oversized postings. 


\mysubsection{Reassign}
A reassign job is generated by merge or split jobs. It checks if vectors in the new postings and/or their neighbors need to be relocated to re-balance the data distributions in the local region. The reassignment check is based on the two \emph{necessary conditions} in \S\ref{ssec:reassign}. 
\revisionstart Note that neighbor posting check is \emph{not} required for merge-triggered reassign. \revisionend


Reassigning a vector without deleting \revisionstart its replicas \revisionend in \revisionstart the unexamined postings \revisionend increases the replica number. 
This not only increases storage overheads but also increases split and reassign frequency since the extraneous replicas take up spaces of postings. 
In order to efficiently identify stale vectors after reassignment without actual deletes, \rebuilder uses a version map to record \emph{the version number} for each vector. A version number takes one byte and is stored in memory to record the version changes of a vector: seven bits for re-assign version and one bit for deletion label.
When reassigning a vector, we increase its version number in the version map and append the raw vector data with its new version number to the target posting.
All the old replicas with a stale version number are dropped during the search. The replicas will be garbage collected later. 

\subsubsection{Concurrent Rebuild}


In \OurSys{}, \rebuilder is multi-threaded with efficient concurrency control of updates to the in-memory and on-disk data structures. Concurrent rebuild can avoid \revisionstart drops \revisionend of index quality due to slow re-balance.

\mysubsection{Concurrency Control for Append/Split/Merge}
Since append, split, and merge may update the same posting and the in-memory block mapping concurrently, 
We add a fine-grained posting-level write lock between these three operations to ensure a posting \revisionstart change \revisionend is atomic. 

Posting read does not require a lock. Therefore, \revisionstart identifying vectors for reassignment is \revisionend lock-free since it only searches the index and checks the two necessary conditions. 
\revisionstart Our experiments show that \revisionend even in a skewed workload, write lock contention is low, i.e., less than 1\% contention cases. This is because only a small portion of postings are being edited concurrently. 

During a reassign process, it is possible that a vector \revisionstart appends to a stale posting, which happened to be deleted concurrently. \revisionend In such a case, we abort the reassignment and re-execute the reassign job for this vector. 
In our experiments, there are only less than 0.001\% of total insertion requests encountering the posting-missing problem caused by split. As a result, the abort and re-execution overhead is minor. 

\mysubsection{Concurrent Reassign}
\revisionstart \OurSys{} avoids \revisionend concurrently reassigning the same vector at the same time.  
When collecting vector candidates for reassignment, \revisionstart \rebuilder \revisionend gathers the current version of the candidates. \rebuilder atomically executes reassignment operations by leveraging atomic primitives of compare-and-swapping (CAS) for the version number.
If an atomic CAS operation fails on the vector version map, reassignment is aborted since the vector becomes a stale version. 
Otherwise, we let the corresponding reassign proceed to the end. 

\begin{figure}[t!]
\includegraphics[width=\columnwidth]{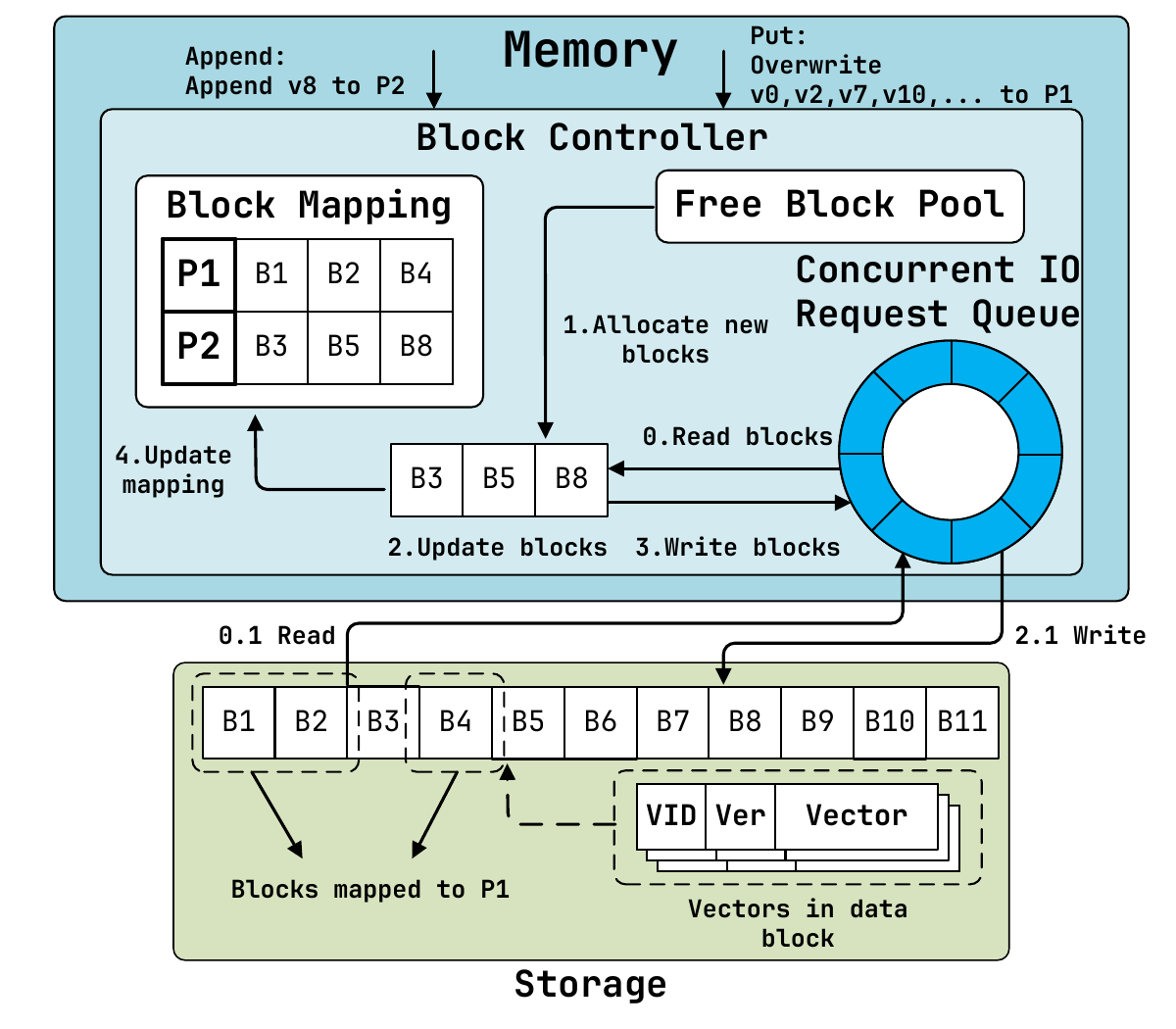}
\caption{\OurSys{} \revisionstart storage overview. \revisionend }
\label{fig:stroagearchitectur4e}
\end{figure}

\subsection{\blockctrl Design}
\label{sect:imple:storage}
\blockctrl is a light-weight storage engine highly optimized for reading. It offers \emph{append-only} operation on postings.
This design 
takes advantage of the characteristics of postings, where old posting data is immutable, and the update introduces no additional overheads for reading (unlike a log-structured file system, where multiple additional out-of-place reads are required). It keeps appending vector updates to a posting before exceeding a length limit. 
When a posting exceeds the length limit, the old posting is destroyed after being split into two new ones.
To avoid unnecessary overheads in file systems or other storage engines (e.g., KV store), \blockctrl operates directly on raw SSD interfaces.

\revisionstart \mysubsection{SPDK-based Implementation} \revisionend
Figure~\ref{fig:stroagearchitecture} overviews the storage design.
\blockctrl is implemented on top of SSD. 
It leverages the raw \emph{block} interface offered by SPDK~\cite{SPDK}, a high-performance NVMe SSD library by Intel. SPDK offers a set of user-space IO libraries for accessing high-speed NVMe devices, which allow us to bypass the legacy storage stack to perform SSD I/Os directly. 

\revisionstart \mysubsection{Storage Data Layout} \revisionend
As shown in Figure~\ref{fig:stroagearchitecture}, a \blockctrl consists of in-memory \emph{Block Mapping}, a \emph{Free Block Pool}, and a \emph{Concurrent I/O Request Queue}.

\textit{Block Mapping} maps a posting ID to \revisionstart its SSD block offsets\revisionend. 
Since the posting ID is a continuous integer, block mapping is implemented as an in-memory dense array, where each element stores the block metadata of a posting length and its SSD block offsets. \revisionstart A posting consists of a list of tuples in the form of <vector id, version number, raw vector>, which typically takes three to four SSD blocks. A block mapping entry only consumes 40 bytes of memory. For one billion vectors, there only exist 0.1 billion postings. In this case, block mapping only consumes about 4GB of memory. \revisionend 

\textit{Free Block Pool} maintains all free SSD blocks. It keeps track of the offsets of all the free blocks to serve disk allocation, and garbage collects stale blocks after spilt and reassign.

\textit{Concurrent I/O Request Queue} is implemented using an SPDK circular buffer, which sends asynchronous read and write to SSD device for maximized IO throughput and low I/O latency. 

\mysubsection{Posting API \& Implementation}
\revisionstart \blockctrl provides \revisionend a set of posting APIs as follows: 
\begin{compactitem}[\noindent$\bullet$]
    \item \verb|GET| retrieves posting data by the given ID. The request first looks up the block mapping to identify the corresponding SSD blocks. Asynchronous I/Os are then sent to the current I/O Request Queue. Later, all desired blocks are collected upon the completion of all I/Os.
    \item \verb|ParallelGET| reads multiple postings in parallel to amortize the latency of individual \verb|GET|s. This ensures fast search and update. 
    \verb|ParallelGET| allows sending a batch of I/O requests to fetch all the candidate postings, which hides the I/O latency and boosts disk utilization.
    \item \verb|APPEND| adds a new vector to a posting's tail. 
    Instead of read-modify-write at the posting-granularity, \verb|APPEND| only involves read-modify-write of the last \emph{block} of a posting, which reduces the amount of read/write amplification significantly. 
    As shown in Figure~\ref{fig:stroagearchitecture}, \verb|APPEND| first allocates a new block, reads the original last block if the last block is not full, appends new values to the values from the last block, and then writes it as a new block. After a new block is written, it atomically updates the corresponding in-memory \emph{Block Mapping} entry via a compare-and-swap operation to reflect the change. 
    The old block will be released to the free block pool for later usage. 
    
    \item \verb|PUT| writes a new posting to SSD. Like \verb|APPEND|, it allocates new blocks and writes for the entire posting blocks in bulk. Then it atomically updates the \emph{Block Mapping} entry. If \verb|PUT| overwrites an old posting, it releases old blocks to the \emph{Free Block Pool}.
\end{compactitem}

\blockctrl provides a common abstraction and implementation, which can be generalized for other read-intensive applications (such as widely-used inverted index for search engine~\cite{lucene, elastic}).

\subsection{Crash Recovery}
\OurSys adopts a simple crash recovery solution, which combines snapshot and write-ahead log (WAL). 
Specifically, an index snapshot is taken periodically, and all update requests between adjacent snapshots are collected into a WAL so that a crash can be recovered from the latest snapshot, followed by replaying the WAL. The WAL will be deleted when a new snapshot is generated.

To take a snapshot for a vector index, we need to record both the in-memory and on-disk data structures.
For in-memory index data, we create snapshots for centroid index, version maps in \textit{Updater}, and block mapping and block pool in \textit{Block Controller}, and flush the snapshots to disk. Snapshots are relatively cheap because these data structures take only 40GB for billion-level dataset, which costs 2\textasciitilde3 seconds for a full flush on a PCI-e based NVM SSD. 
For disk data, thanks to our block-level copy-on-write mechanism, we can collect all the released blocks during two snapshots into a \emph{pre-release} buffer, which will be added to the \emph{Free Block Pool} after the next snapshot is recorded. 
Thus all the data blocks modified in the interim can be rolled back to be consistent with the previous snapshot. 
This solution saves a large amount of disk space since we only delay the space release for old blocks during two snapshots. 



\section{Evaluation}
\label{sect:eval}
\revisionstart
In this section, we conduct experiments to answer the following questions:
\begin{compactitem}[\noindent $\bullet$]
\item How does \OurSys{} compare with state-of-the-art baselines in terms of performances, search accuracy and resource usage? (\S\ref{sec:eval_overall})
\item What is the maximum performance of \OurSys{}? (\S\ref{sec:eval_stress})
\item Can \OurSys{} solve the data shifting problem illustrated in \autoref{fig:Motivation}? (\S\ref{sec:eval_datashifting})
\item How to properly configure \OurSys{}? (\S\ref{sec:eval_parameter})
\end{compactitem}
\revisionend

\subsection{Experimental Setup}
\label{sec:experiment_setup}
\mysubsection{Platform}
All experiments run on an Azure lsv3~\cite{Azurelsv3} VM instance, which is a storage-optimized virtual machine with locally attached high-performance NVMe storage.
In particular, we configured the VM with 16 vCPUs from a hyperthreaded Intel Xeon Platinum 8370C (Ice Lake) processor and 128GB memory for our experiments. 

\mysubsection{Datasets}
We use two widely-used vector datasets to evaluate \OurSys:
\begin{compactitem}[\noindent$\bullet$]
    \item SIFT1B \cite{Spacev} is a classical image vector dataset for evaluating the performance of ANNS algorithms that support large-scale vector search. It contains one billion of 128-dimensional byte vectors as the base set and 10,000 query vectors as the test set.
    \item SPACEV1B \cite{Sift} is a dataset derived from production data from commercial search engines. It represents a different form of vector encoding: deep natural language encoding. It contains one billion of 100-dimensional byte vectors as a base set and 29,316 query vectors as the test set.
\end{compactitem}

\mysubsection{Baselines}
We compare \OurSys with two baselines:
\begin{compactitem}[\noindent$\bullet$]
\item \emph{DiskANN} is the state-of-the-art disk-based fresh ANNS system~\cite{FreshDiskANN2021}. It is based on a graph ANNS index and uses an out-of-place update solution.
We configure DiskANN with the same settings as in their paper~\cite{FreshDiskANN2021}. 
For update configurations, DiskANN baseline processes \emph{streamingMerge}, a lightweight global graph rebuild, for every new 30M vectors, where graph degree $R$ equals 64 and insert candidate list equals 75. 
For search configurations, DiskANN baseline uses the default setting with beamwidth equal to 2 and search candidate list L equal to 40 for recall10@10.
\item \emph{SPANN+}, a modified version of SPANN~\cite{ChenW21} which appends updates locally to a posting \emph{without splitting and reassigning}. This is an \emph{append-only} version of \OurSys without the \rebuilder module.
\end{compactitem}


\mysubsection{Workloads}
Three workloads are used in the experiments.
\begin{compactitem}[\noindent$\bullet$]
\item \emph{Workload A} simulates a realistic vector update scenario with 100 million scale of SPACEV vectors. The reason to reduce the scale from 1 billion to 100 million is that DiskANN requires several TBs of DRAM to run 1-billion scale fresh updates (shown in Table~\ref{tbl:rebuild_cost}), which exceeds our machine's capacity. In particular, workload A simulates 1\% update daily over 100 days. To generate updates realistically, we extract two disjoint SPACEV 100M datasets from SPACEV1B, where one is used as the base ANNS index data-set and the other as the update candidate pool.
Each daily update epoch deletes 1\% of vectors randomly from the base ANNS index, and inserts 1\% of vectors randomly selected from the update data pool to the base index.

\item \emph{Workload B} has the same data scale and sampling method as Workload A but with a 100 million scale of the SIFT vector dataset.

\item \emph{Workload C} scales up our experiment data-set to be \emph{billion-scale} using \revisionstart both \revisionend the SIFT dataset \revisionstart and the SPACEV dataset. \revisionend This workload aims at stress testing \OurSys, also with a 1\% daily update rate. 
\end{compactitem}

\mysubsection{Metrics}
\OurSys is designed for online ANNS streaming scenarios. Thus, our evaluation focuses on the following four categories of metrics. 
\begin{compactitem}[\noindent$\bullet$]
    \item Search Performance: We measure tail (\revisionstart P90, P95, P99, and \revisionend P99.9) latency and query per second (QPS) throughput. In particular, we have a hard cut of 10ms for \OurSys and all baselines, where the system finishes the result immediately and returns the current search results.
    \item Search Accuracy: We use the percentage of ground truths recalled by \OurSys{} system to measure accuracy.
    \item Update Performance: insertion and deletion throughputs.
    \item Resource Usages: the memory and CPU consumption.
\end{compactitem}

\subsection{Real-World Update Simulation}
\label{sec:eval_overall}
In this experiment, we compare all the metrics of \OurSys{} with all the baselines on the real-world situation.
We use Workload A and B (see \S\ref{sec:experiment_setup}) to simulate 100 days of real-world updates, and we show that \OurSys{} outperforms baselines in all evaluation metrics.

\begin{table}[!h]
    \centering
    \begin{tabular}{|c|c|c|c|c|}
    \hline
     & DiskANN & SPANN+ & \OurSys \\
     \hline
      Insert & 3 & 1 & 1 \\
     \hline
     delete & 1 & 1 & 1 \\
     \hline
     Search & 2 & 2 & 2 \\
     \hline
     Background & 10 & 2 & 2 \\
     \hline
    Total & 16 & 6 & 6 \\
     \hline
    \end{tabular}
    \caption{Threads allocation for overall performance.}
    \label{Eval:ThreadsAllocation}
\end{table}

\begin{figure*}[!h]
\includegraphics[scale=1]{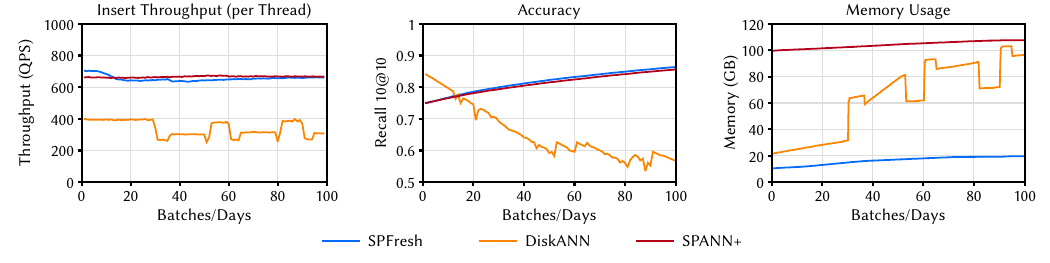}
\includegraphics[scale=1]{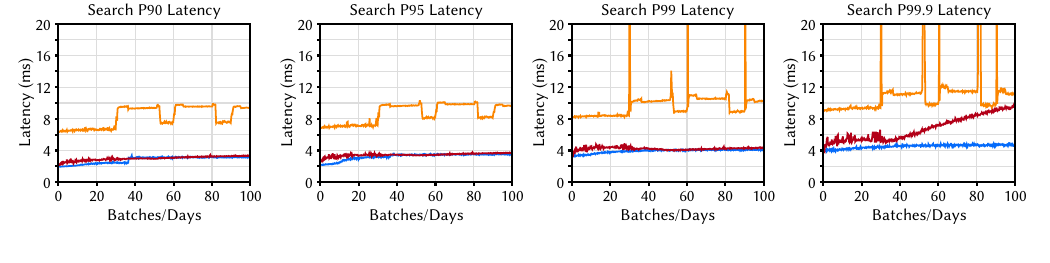}
\caption{\OurSys{} \revisionstart overall performance (SPACEV: data distribution shifts over time). \revisionend }
\label{fig:OverallPerformance1SPACEV}
\end{figure*}

\subsubsection{System Setup}
Table~\ref{Eval:ThreadsAllocation} lists the thread allocation for each system. 
Specifically, we allocate threads to each system's sub-components to meet the processing requirements of handling update throughput of 600\textasciitilde1200 QPS. The setting of update QPS is based on Alibaba's daily update speed (100 million each day)~\cite{ADBV2020VLDB}.
Each system only needs one thread to serve delete requests because deletion uses tombstones to record the deletions, which is lightweight. 
For DiskANN, due to its high insert latency, we set its number of insert threads to 3 and the number of background merge threads to 10, \revisionstart because it is the minimum to keep up with the update and garbage collection process. Further increasing the number will impact the query performance in the foreground negatively. \revisionend 
The two remaining threads are used for foreground search for DiskANN. 

To be comparable with DiskANN, \OurSys{} and SPANN+ also set 2 search threads. 
Both SPANN+ and \OurSys{} only need one insert thread to serve 600+ QPS insert throughput and two background threads for SPDK I/O and garbage collection (or local rebuilder).



\





\subsubsection{Experiment Results}
Figure~\ref{fig:OverallPerformance1SPACEV} records a daily time series of the search tail latency, insert throughput per thread, search accuracy, and memory usage of Workload A. We can see that \OurSys achieves the best and the most stable performance on all the metrics during the 100 days. 

\mysubsection{Low and Stable Search Tail Latency}
\revisionstart Figure~\ref{fig:OverallPerformance1SPACEV} shows that \OurSys{} achieves low and stable tail latency in all percentile measurements.
Since the overall tail latency trends are similar, we focus our discussion on the most stringent tail latency measurement, P99.9.

Experiments indicate that \uproto{} is able to keep posting distribution uniform because \OurSys{} \revisionend has a stable low search \revisionstart P99.9 \revisionend latency around 4ms.
\revisionstart In comparision, \revisionend other systems' \revisionstart P99.9 \revisionend latency is both worse and less stable than \OurSys.

DiskANN's \revisionstart P99.9 \revisionend latency fluctuates significantly with a dramatic increase to more than 20ms during global rebuilds because a search thread could be blocked by a global rebuild even with 10ms hard latency cut. 
The search \revisionstart P99.9 \revisionend latency of SPANN+ increases significantly from 4ms to more than 10ms because its posting keeps growing, inducing data skews and the increase of I/O and computation cost. 

Overall, \OurSys{} maintains 2.41x lower \revisionstart P99.9 \revisionend latency to DiskANN on average and expands its latency advantage to SPANN+. 
\revisionstart
The low and stable search tail latency can be attributed to the LIRE protocol. During the experiment, we found that only 0.4\% insertion will cause rebalancing. Among them, the average split number is 2, and the maximum split number is 160, with a cascading length of 3. The merge frequency is only 0.1\% of the update (insertion and deletion). On average, each time 5094 vectors are evaluated, and only 79 are actually reassigned.
\revisionend

\mysubsection{High Search Accuracy}
\OurSys achieves a higher search accuracy compared to baselines. 
Both SPANN+ and \OurSys{} would not violate the NPA of a \revisionstart cluster\revisionend-based index. Therefore, the accuracy of SPANN+ and \OurSys{} grows gradually since newly inserted vectors are all assigned to a subset of postings due to the data shifting. 
Therefore, queries to these new vectors can easily hit since search and insertion follow the same search path to get the nearest postings. 

Although SPANN+ search accuracy increases in a similar trend like \OurSys, the gap in accuracy increases over time. The increasing gap between these two systems is because the index quality of SPANN+ degrades as partition distributions skew over time.

\revisionstart
DiskANN proposes an algorithm to reduce the overhead of global rebuild by eliminating outdated edges from all vertices and populating edges for a new vertex using the neighborhood of its deleted neighbors.
This method aims to reduce the decline in accuracy due to a decreased number of edges caused by vector deletions without reconstructing its graph-based index completely.
However, experiments show that such a method cannot prevent DiskANN's search accuracy from decreasing over time. 
\revisionend

\mysubsection{High and Stable Update Performance}
\OurSys{} achieves 1.5ms average insert latency and stable tail latency. On the other hand, DiskANN suffers from the heavy computation caused by in-memory graph traversal, and thus results in higher latency and lower throughput. Compared to SPANN+, we can see that \OurSys{}'s lightweight \rebuilder will not affect the foreground insert performance.

\mysubsection{Low Resource Utilization}
For resource usage, \OurSys{} achieves as low as 5.30X lower memory usage than baselines during the whole update process.
DiskANN occupies an extra 60G memories for background streamingMerge and 15GB for the second in-memory index for the update. SPANN+ needs much larger block-mapping entries to allow a larger posting length. \OurSys{} keeps the memory under 20GB, which only grows slightly over time because new metadata are created for each new posting triggered by splits. \revisionstart\OurSys{} also maintains a reasonable disk size. In the index, we find that 86\% of the total vectors have more than one replica, and on average, one vector has 5.47 replicas, which is similar to the index built statically.\revisionend

We also ran the same experiment on Workload B and reached a similar conclusion with DiskANN. Note that SPANN+ achieves similar performance with \OurSys{} on the SIFT dataset, which is almost uniformly distributed. This is expected because background garbage collection should be able to prune stale vectors on SPANN+ without splits on uniform data-set. Consequently, SPANN+ achieves a similar index quality as that of \OurSys{} because its posting distribution does not shift much. 

\revisionstart
\subsection{Billion-Scale Stress Test } 
\revisionend
\label{sec:eval_stress}

We scale up vector data size to billion-level and configure the system to show the best performance \OurSys{} achieves with the given resource. We use Workload C (see \S\ref{sec:experiment_setup})  to simulate a 20-day real-world update scenario. We demonstrate that \OurSys{} has fully saturated SSD's bandwidth and performed well with stable resource utilization.


\begin{figure} [!t]
\includegraphics[scale=1]{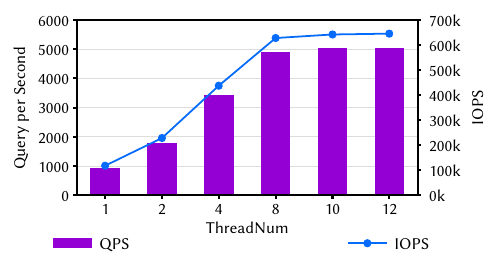}
\caption{Search throughput/disk IOPS vs \# of \OurSys{} search threads on Azure lsv instance~\cite{Azurelsv3}.}
\label{fig:IOPSLimit}
\end{figure}

\begin{figure*}[!ht]
\includegraphics[scale=1]{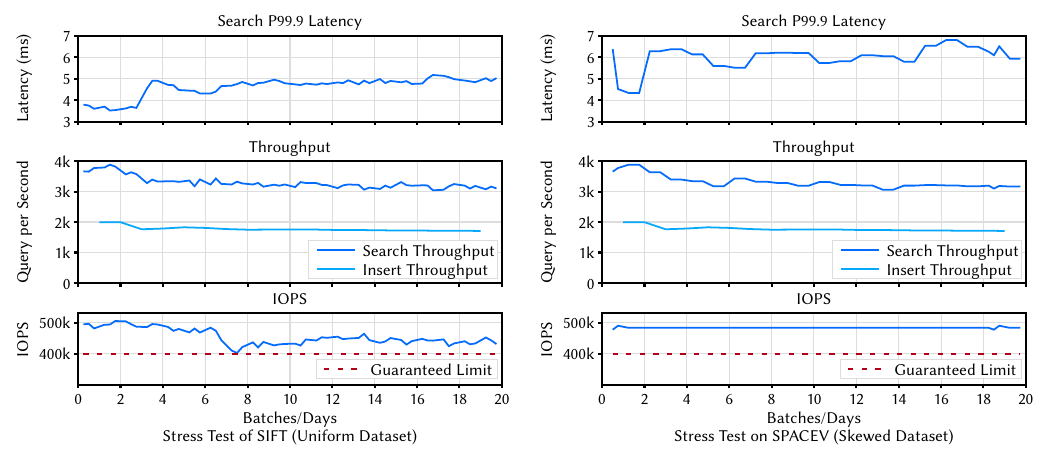}
\caption{\revisionstart Billion-scale \revisionend stress test \revisionstart on uniform and skew datasets. \revisionend \OurSys{} saturates the I/O with a stable \revisionstart P99.9 \revisionend latency while keeping accuracy stable and higher than 0.862 \revisionstart for uniform dataset and 0.807 for skew dataset \revisionend by searching nearest 64 postings, the memory and CPU utilization keep in about 74 GB and 1300\% in \revisionstart both uniform and skew datasets. \revisionend}
\label{fig:StressTest}
\end{figure*}

\subsubsection{System Setup}


\begin{table} [!t]
    \centering
    \begin{tabular}{|c|c||c|c|}
    \hline
     & \#threads & & \#threads\\
     \hline
      Delete/Re-insert & 4 & Search & 8 \\
     \hline
     Background & 3 & Total & 15 \\
     \hline
    \end{tabular}
    \caption{Thread allocation for \OurSys{} in \revisionstart billion-scale tests. \revisionend }
    \label{Eval:ThreadsAllocationStressTest}
\end{table}

Table \ref{Eval:ThreadsAllocationStressTest} lists thread allocation for \OurSys{}'s stress tests. To fully achieve the IOPS of SSD, our setting maximizes search throughput while supporting maximum update throughput. 

Azure lsv3 has a max \emph{guaranteed} NVMe IOPS of 400K~\cite{Azurelsv3}. We first run an experiment to find out the max search throughput lsv3 can handle. As Figure~\ref{fig:IOPSLimit} shows, the IOPS and search throughput almost reach the peak at 8 search threads on a single SSD disk. 

When search thread count is set to 8 for max search throughput, a fore-ground thread count of 4 saturates the update throughput. Therefore we set the thread counts as in Table~\ref{Eval:ThreadsAllocationStressTest} 
for this experiment.
Search is the most important part of ANNS service, so the stress test we will maximize our search throughput while support the maximum update throughput. Since Azure document\cite{Azurelsv3} note that the max NVMe disk throughput is 400K and can go higher but not be guaranteed to keep the IOPS higher than 400K, we make a simple test on lsv3 NVMe SSD by running \OurSys{} Search On Azure lsv3 to find out that the MAX performance of lsv3 NVMe SSD. and we can see on figure \ref{fig:IOPSLimit} that when we set search thread to 8, the QPS of \OurSys{} and the IOPS of Disk will no more grows, so in stress test we will set search thread to 8 and then maximize the update throughput, and we find out that when fore-ground thread set to 4 the IOPS will reach the peak during the update.

\subsubsection{Experiment Results}
Figure \ref{fig:StressTest} records a daily time series of the search \revisionstart P99.9 \revisionend latency \revisionstart on both uniform and skew datasets, \revisionend search/insert throughput, and the IOPS of WorkLoad C. We can see that \OurSys{} reaches the IOPS limitation with stable performance and resource utilization throughout the entire run.

\mysubsection{High NVMe SSD IOPS Utilization} 
As we can see from \autoref{fig:StressTest}, \OurSys{} always fully utilizes the NVMe's bandwidth, even exceeding the max \emph{guaranteed} IOPS of Azurelsv3.
Thanks to LIRE's lightweight protocol, we can see \OurSys{}'s bottleneck is in the disk IOPS, before reaching the CPU and memory resource limit.

\mysubsection{Stable Search and Update Performance} 
As the data scale grows from 100M to 1B, the search latency is stable, just like that in \S\ref{sec:eval_overall}.
There is some slight increase of \revisionstart P99.9 \revisionend latency in the beginning when the first split jobs are triggered. 
In this case, the \revisionstart P99.9 \revisionend latency increases slightly because of the gradual growth of in-memory index size, which makes the in-memory computation more costly over time.

\mysubsection{Stable Accuracy} During the entire stress test, the accuracy of \OurSys{} remains stable, which is higher than 0.862 \revisionstart for the uniform dataset and 0.807 for the skew dataset \revisionend by searching the nearest 64 postings.

\begin{figure}[!t]
\includegraphics[scale=1]{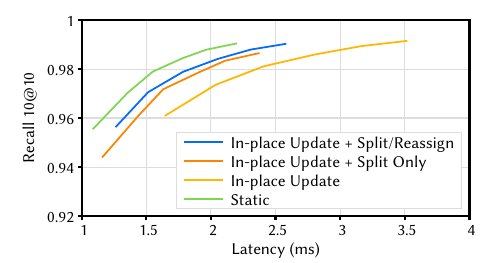}
\caption{The trade-off on search accuracy and latency with various update techniques under an increasingly skewed data distribution.}
\label{fig:DataShifting}
\end{figure}
\subsection{Data Distribution Shifting Micro-benchmark}
\label{sec:eval_datashifting}
In this experiment, we replay the experiment in \S\ref{sect:bg:challenges} to demonstrate that LIRE is required to re-balance the shifting data distribution. 
We compare four systems in this experiment, where \emph{Static} is our target since it has no updates.
For the rest of the three systems, we start with a naive system with in-place update only, i.e., SPANN+, and gradually add sub-components of LIRE into the system.

\mysubsection{Experiment Results}
Figure~\ref{fig:DataShifting} shows the recall and latency trade-off result. With a relaxed search latency, the figure shows that recall improves for all four systems. Meanwhile, as the curve moves northwest, the system shows a higher ANNS index quality with a more accurate recall and a lower latency.
An in-place update-only solution may have a high recall but at the expense of high latency.
Adding a split component into the in-place update decreases the search latency with the same accuracy. 
Adding the reassignment component further decreases the search latency.
As shown in Figure~\ref{fig:DataShifting}, the performance of \OurSys{} with in-place update + split/reassign is the closest one to the Static's results, which represent the ideal cases.


\balance

\subsection{Parameter Study}
\label{sec:eval_parameter}
\revisionstart
In this experiment, we investigate the proper parameter configurations for \OurSys{} to achieve maximum performance. Experiment results show that the Reassignment only requires checking a limited scope of proximate postings for scanning to attain a good index quality. Furthermore, \OurSys{} demands a minimal increase in computational resources, specifically in terms of threads, to accomplish high throughput while also demonstrating good scalability.

\revisionend
\mysubsection{Reassign Range}
The first parameter we examine is the reassign range, i.e., the size of the local rebuild range.
Reassign range is measured by the number of nearby postings to check for vector reassignment after a new posting list is created.
In this experiment, we use the same setting as in \S\ref{sect:bg:challenges}.

\begin{figure}[t]
\includegraphics[scale=1]{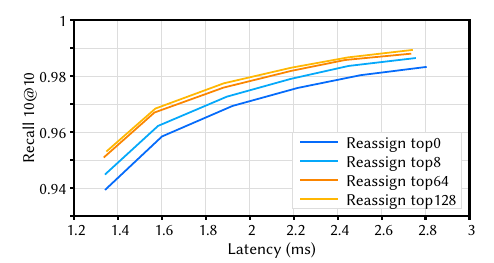}
\caption{Parameter Study: Reassign Range, top64 (nearest 64 postings) is enough for reassign scanning.}
\label{fig:ParameterStudy2}
\end{figure}

In Figure~\ref{fig:ParameterStudy2}, we vary reassign range from the nearest 0, i.e., only process reassign in the split posting, to the nearest 128 postings. As the reassign range increases, accuracy also increases with the same search time budget because more NPA-violating vectors are identified and reassigned.
The accuracy increase rate wanes off as the reassign range increases, where there is only a marginal increase from range 64 to 128.
Consequently, we chose 64 as \OurSys{}'s default reassign range. 

\mysubsection{Fore/Back-ground Update Resource Balance}
The foreground \updater and background \rebuilder work as a feed-forward pipeline as detailed in \S\ref{sect:imple:rebuilder}. In this experiment, we examine the proper resource ratio for \updater and \rebuilder to make their processing speed balanced in the pipeline. Specifically, we configure the foreground and background threads and measure the throughput to see when the update resource is balanced.

\begin{figure}[h]
\includegraphics[scale=1]{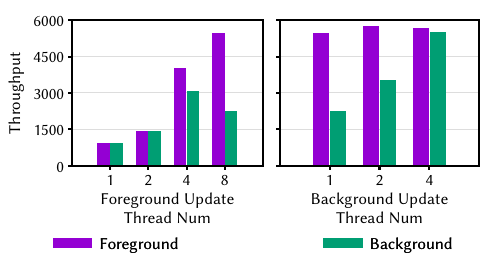}
\caption{Foreground Scalability (Background Thread Number = 1) and Background Scalability (Foreground Thread Number = 8).}
\label{fig:Scalability1}
\end{figure}

As the left part of Figure~\ref{fig:Scalability1} shows, a single-threaded background \rebuilder can keep up the foreground \updater until foreground threads are set to 2. 
Similarly, as on the right side of Figure~\ref{fig:Scalability1}, an 8-threaded foreground \updater needs at least four threads for foreground \updater to generate enough requests for the \rebuilder. Based on the result, \OurSys{} sets a thread ratio of 2:1 between the foreground \updater and the background \rebuilder balances the feed-forward pipeline.


\section{Conclusion}
\OurSys{} supports incremental in-place update for billion-scale vector search. It implements \uproto{}, a Lightweight Incremental RE-balancing protocol to \emph{split} overly large postings and \emph{reassign} vectors across neighboring postings when necessary. Experiments show that \OurSys{} can incorporate continuous updates faster with significantly lower resources than existing solutions while maintaining high search recalls by (1) \uproto{} identifies a minimal set of neighborhood vectors in the large index space for updating to adapt to data distribution shift; (2) the index re-balancing operations and the foreground queries are decoupled, and handled by efficient concurrency control mechanisms, avoiding operation interference. \revisionstart SPFresh’s solid single-node performance builds a strong foundation for the future distributed version. \revisionend

\begin{acks}
\noindent We thank all the anonymous reviewers for their insightful feedback, and our shepherd, Nitin Agrawal, for his guidance during the preparation of our camera-ready submission. This work is supported in part by the National Natural Science Foundation of China under Grant No.: 62141216, 62172382 and 61832011, and the University Synergy Innovation Program of Anhui Province under Grant No.: GXXT-2022-045. Cheng Li and Qi Chen are the corresponding authors.
\end{acks}

\balance

\newpage

\bibliographystyle{ACM-Reference-Format}
\bibliography{spfresh}


\end{document}